# Mechanism of Electrolyte-Induced Brightening in Single-Wall Carbon Nanotubes.

Juan G. Duque,[†] Laura Oudjedi,[‡,#] Jared J. Crochet,[†] Sergei Tretiak,[§] Brahim Lounis,[‡,#] Stephen K. Doorn[§,*] and Laurent Cognet,[‡,#,*].

[†]Chemistry Division, Physical Chemistry and Applied Spectroscopy Group, MS J536, Los Alamos National Laboratory, Los Alamos, NM 87545, [‡]University of Bordeaux, LP2N, 33405 Talence, France, [#]CNRS & Institut d'Optique, LP2N, 33405 Talence, France, [§]Center for Integrated Nanotechnologies, MS K771, Los Alamos National Laboratory, Los Alamos, NM 87545.

*Supporting Information Placeholder*

**ABSTRACT:** While addition of electrolyte to sodium dodecyl sulfate suspensions of single-wall carbon nanotubes has been demonstrated to result in significant brightening of the nanotube photoluminescence (PL), the brightening mechanism has remained unresolved. Here, we probe this mechanism using time-resolved PL decay measurements. We find PL decay times increase by a factor of 2 on addition of CsCl as the electrolyte. Such an increase directly parallels an observed near-doubling of PL intensity, indicating the brightening results primarily from changes in non-radiative decay rates associated with exciton diffusion to quenching sites. Our findings indicate a reduced number of these sites results from electrolyte-induced reorientation of the surfactant surface structure that partially removes pockets of water from the tube surface where excitons can dissociate, and thus underscores the contribution of interfacial water in exciton recombination processes.

Controlling the surface chemistry of nanomaterials is a critical aspect of tuning the electronic and optical properties of a wide range of systems. In the case of single wall carbon nanotubes (SWCNTs), surfactant stabilization is the most common approach to generating the colloidal suspensions of individualized SWCNTs that enable the observation of nanotube PL and study of their photophysical properties.[1] Because all atoms in SWCNTs reside at the surface, their optical properties are highly sensitive to the nature of their surface environment. Choice of surfactant alone or in combination then provides an important route to tailoring the environmental interactions that impact dispersibility and optical properties in various environments.[2-6] A wide range of available surfactants also provides the tunability that has given rise to advances in separation of tubes by chirality and diameter via density gradient approaches.[7] These examples are representative of why significant effort has been made towards understanding and controlling the interactions between dispersing agents and SWCNTs.

As a specific example, sodium dodecylsulfate (SDS)-suspended SWCNTs provide a highly reconfigurable system. Reorganization of the SDS surface structure can be driven by addition of organic solvents[8] or by electrolyte screening.[9,10] Initially, at typical 1% levels, SDS is loosely packed on the tube surface, with relatively random orientation of its hydrophobic chains and anionic headgroups and significant incorporation of water into the surface structure.[11-13] On addition of moderate (tens of mM) levels of electrolyte, headgroup screening results in an increased surface loading of SDS, which drives its reorganization into more densely packed surface structures oriented perpendicular to the tube surface.[9,10] Our earlier work has shown that this reorganization is accompanied by a significant increase in PL intensity, however, the mechanism for this electrolyte-induced brightening remains unresolved.

It has been established that the photophysical response of colloidal SWCNTs is dominated by interactions defined by the surfactant structure and chemistry associated with the nanotube surface.[3,4,14-17] With the SDS-SWCNT system undergoing a transition from a highly disordered to a more well-ordered surface structure on addition of electrolyte, the impact of order on the exciton dynamics may be the underlying source of the brightening. Supporting this assumption, recent molecular dynamics simulations suggest the reordered surface structure may efficiently exclude water from the tube surface[11,13], potentially impacting possible exciton decay pathways. Using PL decay measurements at the single tube level, we have previously shown that the nature of the colloidal interface impacts exciton recombination dynamics and fluorescence quantum yields.[3,14] Similar studies on the

SDS-SWCNT system as electrolyte is added may therefore be a useful probe of the role that surfactant organization plays in determining exciton fate. The local dielectric environment change resulting from surface reorganization may also contribute to the brightening through intrinsic process modifications such as radiative decay. Likewise, by potentially probing impacts on radiative and non-radiative decay pathways, PL decay studies should allow to evaluate the role that changes in the dielectric environment play in the electrolyte-induced brightening.

We focus in this study on changes in exciton dynamics resulting from addition of CsCl, which produced the largest PL quantum yield increase in our previous work.[10] As expected, the increase observed in ensembles of nanotubes is also found across the observable tube population at the single tube level. Figs. 1a and 1b show diffraction-limited images of (6,4) tubes individually dispersed in 1% SDS. Images were collected using the same SWCNT dispersion before (Fig. 1a) and after (Fig. 1b) addition of electrolyte. PL spectra acquired from individual (6,4) tubes (Fig. 1c) reveal the integrated intensity increases by a factor of 1.7 ± 0.2. We note this brightening factor observed at the single tube level closely matches the optimal near-doubling of intensity observed in SWCNT ensembles upon addition of only 10-20mM CsCl.[10] Additionally, the PL spectral peak maximum shifts from 1.401 ± 0.001 eV to 1.404 ± 0.001 eV, while its half-width decreases from 49 to 35 ±3 meV when CsCl is added.

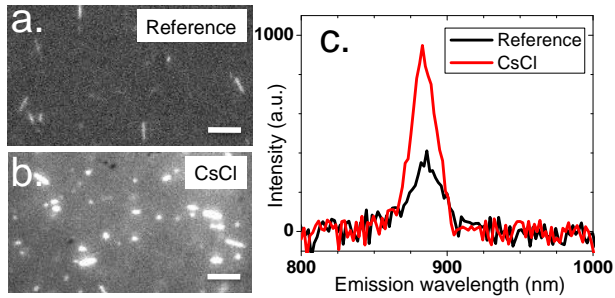

Figure 1. PL images of individual nanotubes suspended in 1% SDS water solution (a) before and (b) after 10 mM CsCl addition. Scale bar 5 μm (c) PL spectra of typical individual (6, 4) tubes before and after CsCl addition.

PL decays for the same SDS-SWCNT dispersions were measured before and after CsCl addition from a near confocal volume using pulsed laser excitation (~ 2 ps pulse-widths) near the (6,4) second-order ($E_{22}$) resonance. A representative series of images (100 ms integration time) taken from the area used to excite the nanotubes and to obtain the PL decay data is shown in Fig. 1S. Multiple tubes diffuse in and out of the excitation volume over the course of our measurement (10 min). The resultant decays thus represent the average behavior of a small ensemble of (6,4) tubes. The PL response is well-fit by a biexponential decay convoluted with the instrument response function of the time correlated single photon counting system (Fig. 2S), consistent with previous work performed on immobilized single nanotubes.[3,14,18] We note $\tau_S$ and $\tau_L$ the short and long components (with amplitudes $A_S$ and $A_L$ respectively) of the biexponential fit: $I(t) = A_S e^{-t/\tau_S} + A_L e^{-t/\tau_L}$. Such bi-exponential behaviors observed in standard samples of micelle encapsulated nanotubes (e.g. deoxycholate (DOC)) were explained previously using a model accounting for the band-edge exciton fine structure which includes the two lowest bright and dark singlet states and the dominant defect-dependent non-radiative decay mechanisms proposed by Pereibenos et al.[19] In particular, we showed previously that high-quality luminescent tubes systematically show biexponential PL decays with $\tau_S$ (tens of ps) reflecting the decay of the bright state and $\tau_L$ (hundreds of ps up to a few ns) reflecting the decay of the dark one, lying a few meV below the bright one. The decay components extracted from the fits for a number of measurements (as exemplified in (Fig. 3S)) are shown in Fig. 2.

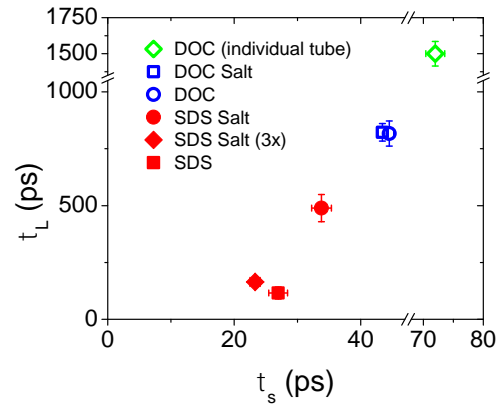

Figure 2. Long vs short PL decay time components obtained from different SWCNT dispersions in 1% SDS and DOC before and after addition of 10 mM CsCl.

Addition of CsCl results in a significant increase in both decay components, with $\tau_S$ increasing from 25 ± 3 to 35 ± 4 ps, and $\tau_L$ from ~100 ± 7 to ~500 ± 34 ps, pointing toward a central role of non-radiative recombinations in this process. We note that the weighting of the long component increases with addition of CsCl and find typical fractional yields $A_L \tau_L / (A_S \tau_S + A_L \tau_L)$ of ~ 2.5 % before salt addition and ~ 4 % after. Increases in both defect density or environmental disorder can affect $\tau_S$ and $\tau_L$.[3,14] Here, because the relative defect density in our samples remains the same across all PL decay measurements, any change in PL decays will reflect solely a change in the local order as driven by electrolyte screening. Thus, the increased decay times observed here suggest that the increased PL intensity following electrolyte addition arises from the SDS reorganization into a more ordered surface structure that ultimately reduces the density of non-radiative decay channels.

This possibility is further supported by the observation that PL decays obtained from identical nanotubes suspended in DOC show longer lifetimes ($\tau_S$ = 45 ± 3 ps and $\tau_L$ = 820 ± 53 ps, Fig. 2) consistent with previous single nanotube observations. The longer decay times can be understood as a result of a more ordered and dynamically stable surface structure available with DOC due to its ability to effectively tile the SWCNT surface via specific interactions of the DOC hexagonal ring structures with those of the SWCNT.[2,4,20] Some variation in surface structure is expected from tube-to-tube within the small ensembles measured here and will be reflected in shorter average decay times from what might arise for a single tube with an optimized surface structure. Such a case may be found at the true single-tube level if DOC-suspended SWCNTs are immobilized in an agarose gel matrix. For such single (6,4) tubes measured here, the longest value for $\tau_S$ (72 ± 8 ps) and $\tau_L$ (1400 ± 243 ps) (Fig. 2)[18] shows an extension of the linear behavior in the decay times that represents a steady progression in response from highly disordered to highly ordered surface structures. We also note that no change in decay times is observed on addition of CsCl to DOC-suspended SWCNTs (Fig. 2), which correlates with the lack of any intensity changes found with electrolyte for DOC suspensions[10] and further highlights the stability of the DOC surface structure. Likewise, electrolyte reorganization of SDS also likely results in a more stable surface structure that is less prone to dynamic fluctuations.

We finally determined the tube PL properties with different salt concentrations: upon 5 mM salt addition, electrolyte reorganization of SDS is already highly efficient (Fig 4S), consistent with previous ensemble experiments[10] and for high electrolyte concentrations (30 mM Cs+), i.e. above the SWCNT aggregation threshold,[9,10] significant loss of PL intensity accompanies aggregation of tubes. In the latter case we find a significant decrease in $\tau_S$ and $\tau_L$ (23 ± 3 and 165 ± 9 ps, respectively) in the same range as found prior to Cs+ addition. This may result from coupling to metallic tubes or from intertube interactions within the SWCNT aggregates acting as an extended defect,[21,22] with both effects acting to increase the number of non-radiative decay pathways.

The above results highlight the strong correlation between surface disorder and exciton recombination dynamics. How this disorder ties to the electrolyte-induced brightening may be determined from a more quantitative analysis of the decay time changes with respect to the parallel changes in PL intensity. Moreover, while the effect of water exclusion due to SDS reorganization on PL intensities and decays is in part accounted for within the concept of ordered vs. disordered environments, the above discussion does not provide an underlying mechanism.

Recently, several groups have shown how exciton recombination rates are linked to the exciton mobility through diffusion limited quenching processes[15-17,23]. Because both $\tau_S$ and $\tau_L$ are modified upon salt addition, non-radiative processes must play a central role in our observations. The effect on radiative processes needs however to be clarified. We now define $\tau_{PL}$ as an effective decay time calculated as a weighted average of the long and short decay components: $\tau_{PL}$ = ($A_S\tau_S$ +$B_L\tau_L$)/($A_S$ +$B_L$). This gives 27 ± 2 ps and 54 ± 3 ps before and after addition of CsCl, respectively, and demonstrates a doubling of the $\tau_{PL}$ upon addition of salt. By next analyzing the fluorescence quantum yield $\eta$ =$\tau_{PL}$ /$\tau_r$ (with $\tau_r$ = radiative lifetime) we can estimate to what extent changes in radiative decay rates might contribute to the brightening mechanism. We find (Fig. 1c) a factor of 1.7 to two-fold increase in fluorescence intensities. From the ratio of quantum yields $\eta_{ns}$ / $\eta_{ws}$ = $\tau_{PL,ns}\tau_{r,ws}$ / ($\tau_{PL,ws}\tau_{r,ns}$), this increase can be completely accounted for by the doubling of $\tau_{PL}$ on addition of CsCl. This indicates that radiative lifetimes likely remain weakly affected by the surfactant reorganization process, i.e. $\tau_{r,ws}$ ≈ $\tau_{r,ns}$ and thus allow investigating the role that water exclusion from the nanotube surface might play in the brightening process.

Radiative decay rates are proportional to exciton oscillator strength, which in turn depends on the dielectric environment through its impact on exciton size (defined here as the separation distance of the bound electron-hole pair).[24] The reduction in polarity accompanying exclusion of water in the reorganized SDS environment could thus be expected to modify the radiative rate on electrolyte addition[25]. For instance, changes of dielectric environments induced by the presence of water inside nanotubes were previously shown to account for increased radiative rates upon insertion of endohedral water inside carbon nanotubes[18]. Here we find that the impact on radiative decay of such a reduction in polarity must only be a minor contribution, as the PL intensity increase can be understood solely in the context of changes in non-radiative decay rates. The exclusion of water thus likely plays more of a role via its contribution to non-radiative decay by providing local structures that act as exciton quenching sites.

Water-mediated exciton quenching can arise from proximity effects of confined water near the nanotube surface. It is known that when water is confined to the core of micelles electrophilic properties can be enhanced relative to bulk water[26]. Similar effects arising from altered hydrogen bonding are expected at the tube surface, as reflected in slower characteristic vicinal water motion relative to the bulk.[27] Moreover, water molecules near the carbon surface can align with the hexagonal carbon networks and have preferred orientations. An increased interfacial dipole moment results,[28] with enhanced dielectric screening providing a route to exciton dissociation[29]. Additionally, the modified water electrophilicity is capable of shifting electron density away from the nanotube surface in the vicinity of local water structures, establishing them as exciton dissociation sites[29]. Without salts, dynamic local water structures will thus act as quenching sites, while the exclusion of water from the nanotube surface on addition of electrolyte will reduce the number of quenching sites and increase PL quantum yield. While these two routes to water-mediated exciton dissociation exist, our observation of only a small blue shift in emission wavelength on addition of electrolyte suggests the electrophilic effects may dominate. Finally, it is important to note that the local water structures we discuss are certainly not static. The more stable surfactant structure formed on electrolyte addition therefore also likely inhibits the dynamic formation of such quenching sites and may be an additional factor in determining the degree of PL brightening.

In conclusion, measurement of PL decays has provided a route to resolving the mechanism of electrolyte-induced brightening of SDS-suspended SWCNTs. Brightening is found to occur predominantly from a reduction in non-radiative decay rates arising from a surfactant reorganization that partially drives out spatially confined water. These findings clarify the role of interfacial water in favoring non-radiative exciton recombination processes and provide further rationale for tailoring SWCNT photophysical response via control over local surface environment.

## ASSOCIATED CONTENT

**Supporting Information**. Materials and methods, and Figures S1–S2. This material is available free of charge via the Internet at http://pubs.acs.org.

## AUTHOR INFORMATION

**Corresponding Author**

*skdoorn@lanl.gov and *lcognet@u-bordeaux1.fr

**Funding Sources**

No competing financial interests have been declared.

## ACKNOWLEDGMENT

This work was supported in part by the LANL-LDRD program and performed in part at the Center for Integrated Nanotechnologies, a U.S. Department of Energy, Office of Basic Energy Sciences user facility. This work was also funded by the Agence Nationale de la Recherche, Region Aquitaine, the French Ministry of Education and Research, and the European Research Council.

Graphic entry for the Table of Contents (TOC)

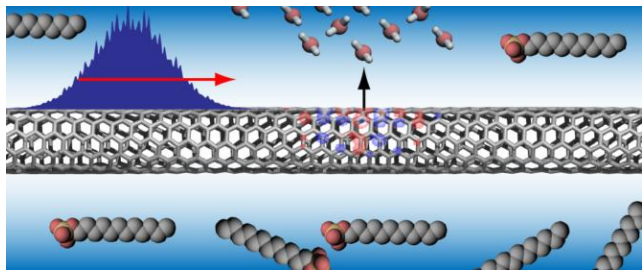